\documentclass[10pt]{article}

\usepackage{latexsym}

\begin{document}

\noindent{\Large \bf
A remark on the notion of robust phase transitions


\normalsize \rm}

\vspace{24pt}

\noindent{\bf  Aernout C.D. van ENTER$^{1}$,

\vspace{12pt}
\it
\noindent
$^{(1)}$ Instituut voor theoretische natuurkunde R.U.G.
Nijenborgh 4, 9747AG, Groningen, the Netherlands.

\rm

\vspace{24pt}
\footnotesize
\begin{quote}
{\bf Abstract:} 
We point out that the high-q Potts model on a regular lattice at its 
transition temperature provides an example of a non-robust -- in the 
sense recently proposed by Pemantle and Steif-- phase transition.

\vspace{3pt}


\end{quote}
\normalsize

\vspace{12pt}



In a recent paper \cite{PS} about phase transitions of spin models 
on general trees, Pemantle and Steif introduced a 
distinction between the notions of  ordinary and "robust" phase transitions.
In their tree calculations, this notion of robust phase transitions
was argued somehow to be more natural. They proposed as an interesting 
question to investigate this distinction also for regular  lattices, and in 
particular the possible occurrence of non-robust phase transitions for spin 
models on regular lattices. 

Here I point out that
at the transition temperature of a high-q Potts model on a regular lattice 
the phase transition is non-robust, as a consequence of the occurrence of 
a first-order transition in the temperature variable. 

A phase transition (for Ising, Potts or n-vector models) is called robust
\cite{PS}, if weakening the bonds at the boundary of some large volume
to an arbitrarily small (but positive) strength in some non-symmetric state
does not influence the 
non-symmetric character of the spin distribution at the origin, in the limit 
where this boundary moves to infinity. 
By this construction one can interpolate between pure and free boundary 
conditions.

In the Ising model on a regular lattice robustness of the phase transition, 
when it occurs, follows from a result of Lebowitz and Penrose\cite{lebpen}, 
that for all temperatures weak positive 
boundary fields in the thermodynamic limit are equivalent to plus boundary
conditions, in the sense that both induce the same  pure (plus) Gibbs measure.

This is in fact not so surprising, as below the critical temperature free (symmetric) boundary conditions 
give rise to a symmetric mixture of pure ordered Gibbs measures, and 
these mixtures are unstable for even a small asymmetry at the boundary. 
After all, there are non-trivial --bounded-- observables at infinity by which 
one can change the symmetric weight distribution of a mixture into an 
asymmetric one, and the interpretation of such observables at infinity as 
uniformly bounded, that 
is, quite weak, boundary terms has been known for a considerable time 
\cite{narrob}, see also \cite{brlepf} for related arguments).

Indeed, let $h$ be a non-trivial observable at infinity, such as exists
for (and only for) a mixed Gibbs measure, say $\mu$, then
\begin{equation}
\mu^{h}(.) = {{\mu(exp \ {h} \times .)}\over {\mu(exp \ {h})}} \neq \mu 
\end{equation}

Thus the Gibbs measure (Gibbsian for the same interaction as $\mu$) 
$\mu^{h}$ which is  obtained formally by adding the bounded term 
$h$ to the Hamiltonian, is different from $\mu$.

Moreover, in the references mentioned above it is described how $\mu^{h}$ 
can be 
approximated by  a sequence of measures of the form $\mu^{h_n}$ where the 
$h_n$ form a uniformly bounded sequence of boundary terms associated to a 
sequence of increasing volumes. 

Thus it is not too 
surprising that  breaking the symmetry in the boundary conditions in the 
manner described before, in which a small strength of a boundary field is 
multiplied by the size of the boundary, has a more drastic effect because 
this sequence of boundary terms 
diverges, instead of staying uniformly bounded. Indeed, in the Ising model,
according to the Lebowitz-Penrose result it immediately drives 
the state to be extremal.

We remark as an aside that the notion of robustness is intimately linked 
with the single-spin space and the Hamiltonian having a symmetry 
which gets broken by the weak boundary field. 

In the high-q Potts case at the transition temperature, however, in contrast to
the low-temperature Ising case, there exists  another
Gibbs measure having the Potts permutation symmetry, 
beyond the symmetric mixture of the ordered measures. This is 
the disordered state, which can be obtained by imposing free boundary 
conditions, and which is pure, moreover\cite{KS,LMMRS,brkule}.

It takes 
more "boundary (free) energy" to move from one  pure state 
to another than from a mixed state to one of its pure components.
Intuitively  put, if one thinks of the pure states as valleys 
in some (free) 
energy landscape, to go from one pure state to another, 
the system has to overcome a free energy barrier of boundary size. 
This represents the cost of inserting a droplet of another (here an ordered)
phase. When the 
boundary bonds are too weak, even if the state outside $\Gamma$ is ordered,
they can't provide sufficient energy to cross this
barrier between disordered and ordered phase, and favor order inside $\Gamma$. 

On the other hand, mixed states are on top of this free energy 
barrier, and they can be easily pushed off from there. 

\smallskip

To be a bit more precise we now adapt the definition of Pemantle and Steif
to a regular lattice. (For the precise definitions, background  and 
general notions of Gibbs measure theory we refer to 
\cite{Geo} or \cite{EFS_JSP}.)

Denote by $\mu_{J,\epsilon,\Gamma}^{+}$ the infinite-volume measure 
which is obtained from the pure plus measure (we call the first of the q 
Potts-states also ``plus'' here by abuse of terminology) by
multiplying the bond strengths $J$ in some contour $\Gamma$ by $\epsilon$.
$\Gamma$ is a set of bonds (or edges) such that their dual bonds (plaquettes)
form a closed (hyper-)surface separating the inside and outside of $\Gamma$.
It plays the role of the ``cutset'' of \cite{PS}. The choice of  the plus
measure induces an effective boundary term, favoring order in  
the plus direction, at the boundary  $\Gamma$ of the volume $Int(\Gamma)$, 
the strength of which is bounded above by $2d \times \epsilon \times \Gamma$. 

\noindent
{\bf Remark:} 

We remind the 
reader that the (infinite-volume) plus measure is non-symmetric, 
due to the assumption of 
existence of a phase transition. Thus we have taken a first thermodynamic limit
already, and (by the DLR-equations) the marginal measure to the 
configuration-space determined by the spins in the interior of $\Gamma$ is 
an average over measures with different boundary conditions, 
all with weak boundary bonds, 
averaged with respect to this  non-symmetric infinite-volume measure 
$\mu_{J,\epsilon,\Gamma}^{+}$. As all boundary bonds in $\Gamma$ are weak, 
this means that one can obtain the same measure by taking a 
suitable  weak boundary term added to the finite-volume Hamiltonian.

\noindent
{\bf Definition}: 

A phase transition of a nearest neighbor Potts or n-vector model is robust if 
for the marginal measure to the single-site space at the origin  
for each positive $\epsilon \in (0,1]$, at least for some subsequence of 
increasing contours $\Gamma_n$ 
whose interiors will finally include each finite volume, 

\begin{equation}
\lim_{\Gamma_n \to \infty} \mu_{J,\epsilon,\Gamma_n}^{+} \neq \mu_{J}^{free}.
\end{equation}

Our above discusion can be summarized in the following

\smallskip 
\noindent
{\bf Theorem}: 

For the the q-state Potts model on $Z^{d}$, d at least 2, 
with q high enough, at the transition temperature the phase transition 
is non-robust.  

\smallskip
\noindent
{\bf Proof}: 

We claim that for small enough $\epsilon$ the above inequality (2)
does not hold. 

There are different ways one might approach a proof. We will sketch here 
how one can adapt the Fortuin-Kasteleyn random-cluster representation 
Pirogov-Sinai contour arguments of \cite {LMMRS} to 
obtain one. In this random-cluster representation (for a detailed description 
of the random-cluster representation for Potts models, see for 
example \cite{FK,edwsok,ACCN,Gri,gehama}),
one considers an associated correlated edge-percolation 
model, in which in a finite volume $\Lambda$ the probability of an 
edge configuration $\eta \in (0,1)^{B(\Lambda)}$ is given by:

\begin{equation}
\mu^{\Lambda} (\eta) = {1 \over Z}{ \prod_{e \in B(\Lambda)} 
p_e^{\eta_e} (1-p_e)^{1- \eta_e} q^{C(\eta)}}, 
\end{equation}

where $C(\eta)$ denotes the number of occupied connected clusters in the 
configuration $\eta$, and

\begin{equation}
p_e = 1 - exp( \ -J_e) \ ,
\end{equation} 

with $J_e$ the bond strength along edge $e$.

Percolation in the random-cluster model occurs if and only if 
there is long-range order in the associated spin model.

First we notice that the finite-volume measures obtained by taking wired 
boundary conditions outside a volume $\Lambda_m$ containing $\Gamma$ 
decrease, in FKG-sense, as the $\Lambda_n$ 
grow, and each of them FKG-dominates the marginal on $int(\Gamma)$ of the
infinite-volume wired measure on  with  ``weak'' bonds in $\Gamma$.
The wired boundary conditions correspond to having all edges occupied outside
the region, or, in spin language, to having all spins aligned (for instance in 
the plus configuration). In particular this infinite-volume wired state
is associated to the measure $\mu^{+}_{J, \epsilon, \Gamma}$. 

These observations have several implications:

\begin{itemize} 

\item[ (i) ] The limit in the left-hand side of (2)) always exists (in the 
weak sense). We emphasize that it is, in fact, a double limit formed by first 
taking for each $\Gamma_n$ wired boundary conditions outside an increasing 
sequence of volumes $\Lambda_m$ containing (the for the moment fixed) 
$\Gamma_n$, (this limit 
exists as it is a limit of FKG-decreasing measures), and then 
taking the limit $\Gamma_n \to \infty$.

\item [ (ii)] If instead we take a ``diagonal'' limit $\Gamma_n \to \infty$ 
with wired boundary conditions outside $\Gamma_n$, we will see that we 
get a convergent 
sequence.  Its limit is a measure that dominates any subsequence 
limit of the left-hand 
side of (2) (as for each $\Gamma$ the element of the sequence dominates the 
corresponding element of the sequence in (i) ). 
It turns out that it approaches the FKG-minimal state, that 
is, (for high q) the disordered state.

\item [ (iii) ] As the limit in (ii) is the FKG-minimal random-cluster state, 
so is the limit in (i).

\end{itemize}

 
   

We can take the $\Gamma$'s to be the boundary of large squares in d=2, 
or (hyper-)cubes in higher dimensions, with the origin at the center.  

In the first version of this paper I sketched how this weakly wired
limit (ii) can be compared with (and can be shown to coincide with)
the limit of measures with free boundary conditions through 
adaptation of existing proofs of the first-order Potts 
transition. I emphasized  the arguments in \cite{brkule} (see Appendix), 
but also mentioned
the random-cluster version. After submitting this version,
R. Koteck\'y kindly informed me that a detailed 
version of  the necessary contour analysis for the random-cluster version, 
essentially along this line, was worked out by I. Medved \cite{med}, and 
included in his more general analysis of finite-size effects. In his 
terminology, having wired boundary conditions outside $\Gamma$  and 
weak enough
bonds in $\Gamma$, is an example of having ``disordering boundary conditions''.
In his Section (2.2) a range of $\epsilon$ which are disordering is determined.
 
With the above FKG-domination-argument, the statement of the 
non-robustness  becomes 
indeed a corollary of his results. $\Box$

\smallskip
Although the proof thus rests on Medved's result, let me add some explanatory 
remarks, see also the Appendix.
The proof goes essentially along the lines of \cite{LMMRS}, once one 
notices that,  because of the above, for small $\epsilon$ only a small fraction
of the boundary edges in $\Gamma$ (in the FK-representation \cite{ACCN,FK,
edwsok,Gri}) are occupied.

Thus the system acts essentially like the system with  
free boundary conditions outside $\Gamma$
(up to a  boundary term of order $\epsilon J$ $\times$ $|\Gamma|$ which is
small compared to the boundary free energy term necessary to induce
the wired state on the inside of $\Gamma$ which is of order $J \times \Gamma$).
Differently put, the ``weakly wired'' boundary conditions do
not influence the behavior in the infinite-volume --- that is now 
infinite-$\Gamma$ --- limit, as compared to the free boundary conditions. 
The reason is that the Peierls contour 
estimate for the probability of finding an essentially 
ordered  region inside $\Gamma$ is exponentially small in $|\Gamma|$.
Multiplying by a term $exp(\epsilon J \Gamma)$ does not qualitatively 
change this. 
At this point it is essential that q is large enough,
and that one is at the transition temperature where the pure disordered state 
coexists with the q ordered ones.
For the low temperature Ising model with 
free boundary conditions outside $\Gamma$ the probability of finding
the system inside $\Gamma$ in an essentially plus configuration is $1 \over 2$,
and similarly for Potts models in the low temperature region this probability
is $1 \over q$. These probabilities follow from symmetry, and are not obtained
by contour estimates. It is here that the essential difference with the 
non-robustness example occurs.


 

For an earlier analysis of a related situation of 
changing ``pure-state'' to ``almost-pure-state'' boundary conditions in 
a more symmetric set-up see \cite {ccf}. 


\bigskip

\smallskip
\noindent
{\bf Comment 1:} 

The mechanism which causes the transition to be non-robust, is the fact
that there is a first-order transition in temperature, such that at the 
transition temperature there is coexistence between a higher-entropy, 
lower-energy disordered
state of higher symmetry, and a number of lower-entropy higher-energy 
states which are of lower symmetry. Thus in more complicated models, 
one expects the transition also to be non-robust whenever 
there is a first order transition in temperature, accompanied by the breaking  
of a symmetry on the low temperature side of the phase transition.

\bigskip

\smallskip 
\noindent
{\bf Comment 2:}

Although also  for the high-q (in fact already for any value of q larger than 
2) Potts model on trees the notions of ordinary and robust phase transitions do
not coincide, and although, moreover, the high-q Potts model on trees, just as 
on regular lattices, typically shows a first-order transition in the 
temperature \cite{PS}, the interpretation of the distinction between ordinary 
and robust transitions seems somewhat different on trees. The separation 
between boundary terms and volume terms is more questionable on trees, so an 
interpretation in terms of a "boundary free energy" analysis does not seem 
possible. Indeed, on trees there can be coexistence of two ordered (the plus 
and minus) states with a disordered one,  even for the Ising model, on a whole 
intermediate temperature interval \cite{CCST,blruza,io}. For the Ising model on
trees, however,  Pemantle and Steif have showed that the occurrence of a phase 
transition and a robust phase transition always coincide. 

\section{APPENDIX}

In this appendix I present a way to get an alternative derivation 
of Medved's result. 
In particular, I sketch  how the arguments of the Pirogov-Sinai
proof by Bricmont, Kuroda and
Lebowitz \cite{brkule}, (BKL) should be adapted to handle the case of weak 
boundary bonds.

We give, as in \cite{brkule}, the argument for d=2. We compare only the weight 
of ``boundary squares'', that is squares inside $\Gamma$ such that 
they touch at least one  (and thus two or three) of the weak bonds, as all 
other computations are unchanged. 
The new element as compared to \cite{brkule} is that now an ordered  
boundary square will be a contour square (in the sense of BKL).
Indeed, call a $E_{4}'$ those sites in $E_4$ (that means, they touch four 
ordered bonds), which 
touch at least one weak bond. Such a site has at least 
${1 \over 4} \times {(1- \epsilon)}$ less 
energy
than it would have if none of the boundary bonds were weak, but still 
contributes zero entropy.
This implies that even with ordered boundary conditions,  an ordered 
boundary square with four $E_4$-sites
has lower free energy compared to a disordered boundary square with four
$E_0$-sites. 

The definition of contours in terms of irregular --or contour--
squares is then the same as in BKL.

Thus BKL eq (3.34), which estimates the partition function in volume $\Lambda$
for all 
configurations compatible with a prescribed configuration of broken and 
unbroken bonds, is replaced by
\begin{equation}
Z(\Lambda|\underline u, \underline b) \leq q^{(E_{0} + E_{4} -E_{4}')} q^{({3 \over 4} + \epsilon)( E_{1}+E_{2}+E_{3} +E_{4}')},
\end{equation} 
with $\Lambda$ equal to $int(\Gamma)$,
and BKL eq (3.35), the inequality for the ratio of partition 
functions with or without the constraint that contour $\gamma$ is present 
in volume $\Lambda$ with boundary condition $\omega$, 
which expresses that the Peierls condition applies,
\begin{equation}
Q(\gamma|\Lambda, \omega) \leq 2^{2 |\gamma|} q^{ - {{C \over 4}|\gamma|}}
\end{equation}
still holds, but with a slightly worse constant C.

(As I already used the symbol $\Gamma$, I slightly changed the notation 
of \cite{brkule} to 
let $\gamma$ denote a contour).

For a central square to be ordered, it needs to be surrounded by a contour, 
thus, at sufficiently high q and small enough (dependent on q) $\epsilon$, 
this is of low probability, uniformly in the size of the enclosing boundary.

\section*{Acknowledgments}
I learned about the notion of robust phase transitions from 
the Kac seminar lectures of Jeff Steif, whom I also thank for providing me 
with the preprint of \cite{PS}. Some useful conversations with Marek 
Biskup are  gratefully acknowledged. I thank M. Biskup, R. Fern\'andez and 
M. Winnink for some advice on the manuscript. I thank R. Koteck\'y for
informing me of the existence of \cite{med}, and explaining to me how the 
necessary estimates are contained in Medved's work and R. Koteck\'y and 
M. Biskup for making a copy available to me,



\end{document}